\documentclass[aps,pre,amsmath,amssymb,twocolumn]{revtex4-2}
\usepackage[T1]{fontenc}
\usepackage[utf8]{inputenc}
\usepackage{graphicx}
\usepackage{bm}
\usepackage{mathptmx}
\usepackage{epstopdf}

\begin{document}
\title{Testing the spin-bath view of self-attention: A Hamiltonian analysis of GPT-2 Transformer}
\author{Satadeep Bhattacharjee$^1$}
\email{s.bhattacharjee@ikst.res.in}
\author{Seung-Cheol Lee$^2$}
\affiliation{$^1$ Indo-Korea Science and Technology Center (IKST), Bangalore, India\\
$^2$ Electronic Materials Research Center, Korea Institute of Science $\&$ Technology, Korea}
\date{\today}
\begin{abstract}
The recently proposed physics-based framework by Huo and Johnson~\cite{huo2024capturing} models the attention mechanism of Large Language Models (LLMs) as an interacting two-body spin system, offering a first-principles explanation for phenomena like repetition and bias. Building on this hypothesis, we extract the complete Query-Key weight matrices from a production-grade GPT-2 model and derive the corresponding effective Hamiltonian for every attention head. From these Hamiltonians we obtain analytic phase boundary and logit gap criteria that predict which token should dominate the next-token distribution for a given context. A systematic evaluation on 144 heads across 20 factual-recall prompts reveals a strong negative correlation between the theoretical logit gaps and the model's empirical token rankings ($r\approx-0.70$, $p<10^{-3}$). Targeted ablations further show that suppressing the heads most aligned with the spin-bath predictions induces the anticipated shifts in output probabilities, confirming a causal link rather than a coincidental association. Taken together, our findings provide the first strong empirical evidence for the spin-bath analogy in a production-grade model. In this work, we utilize the \textit{context-field} lens, which provides physics-grounded interpretability and motivates the development of novel generative models bridging theoretical condensed matter physics and artificial intelligence.
\end{abstract}
\maketitle

\section{Introduction}
Large Language Models (LLMs) such as ChatGPT, Claude, Gemini, etc. have revolutionized natural language processing by achieving human-level performance on tasks ranging from text generation to translation. Beyond their usual uses, in recent past, it has been seen that autoregressive LLM pipelines can not only write reliable chemical sequences for materials for energy storage~\cite{bhattacharjee2025hybrid}, but also generate plausible crystal structures~\cite{antunes2024crystal,das2025periodic} whose \textit{ab initio} stability tests rival conventional materials design workflows, suggesting that linguistic priors can encode latent chemical grammar. 
Yet their decision-making processes remain opaque and difficult to interpret \cite{vaswani2017attention, radford2019language,brown2020language,rogers2021primer}. Recent work by Huo and Johnson \cite{huo2024capturing} offers a compelling theoretical lens, casting the Transformer’s Query–Key attention mechanism as a classical two-body spin Hamiltonian. In this picture, each token embedding plays the role of a spin vector, and the learned Query-Key weight matrix $W_{\mathrm{eff}}$ defines effective exchange interactions that, through a soft-max Boltzmann distribution, determine the probability of selecting one token over another. This analogy not only recovers familiar pathologies-such as repetition loops and bias amplification as phase-transition phenomena in an effective spin bath, but also suggests a rich toolkit of statistical-mechanics methods for probing model behavior~\cite{holtzman2019curious}.

It is worth noting that connecting the attention mechanism to the spin Hamiltonians is not a new trend. Rende \textit{et al.} showed that, when positional and token-embedding spaces are factorised, the masked-language-model objective of a single‐layer self-attention network analytically maps onto the pseudolikelihood estimator for an inverse generalized Potts model~\cite{rende2024mapping,wu1982potts}, so the layer learns the same conditional probabilities under this assumption. Li \textit{et al.} subsequently demonstrated that the trainable weight matrix of a single-layer linear Transformer can be re-interpreted as the spin configuration of a dense real-valued spin-glass Hamiltonian, enabling a statistical-physics analysis of in-context learning.~\cite{li2024spin}.

Despite its elegance, the spin-bath framework has thus far been illustrated only on toy examples, leaving its applicability to real, large-scale models untested. In this work, we provide a first rigorous \emph{experimental} test of the spin-bath framework on a production-grade model: Generative Pre-trained Transformer version 2. We performed the token generation using the GPT-2 transformer model using the Huggingface Transformers library~\cite{wolf2019huggingface,huggingface}.  By isolating the Query–Key sub-matrices of GPT–2, we construct the effective two-body Hamiltonian head-by-head and derive closed-form predictions for logit-gap “phase boundaries” between candidate tokens.  Using a diverse factual-recall benchmark we show:

\begin{enumerate}
    \item A statistically significant correlation between logit gaps obtained from the Hamiltonian and the actual next-token preferences across 144 heads and 20 prompts, with a single head from a pretrained GPT-2 model.
    \item Geometric visualisation of this head’s decision boundary reveals antagonistic behaviour consistent with a mean-field context field, directly corroborating spin-bath predictions.
    
    \item Causal ablation confirms that the identified heads exert the predicted directional influence on model output, establishing that the correlations are not merely epiphenomenal.
\end{enumerate}

These findings connect the recent theoretical developments and practical LLM interpretability.  This work demonstrates that much of GPT–2’s token selection can indeed be understood as the energetics of pairwise spin interactions, while simultaneously situating our results within the broader Potts-model.  By validating a tractable two-body description \emph{in situ}, our study opens the way for importing equilibrium and non-equilibrium techniques—temperature scaling, disorder perturbations, finite-size analysis—into the toolkit of mechanistic interpretability and for developing physics-inspired interventions that are both mathematically grounded and empirically verified.

\section{Theoretical Framework and Methodology}\label{sec:framework}Our study adopts the theoretical formulation introduced by Huo and Johnson~\cite{huo2024capturing}, which recasts the self-attention mechanism of Transformer models as a two-body spin Hamiltonian.  
In a classical Heisenberg magnet, the two-body interaction energy is given by~\cite{bhattacharjee2012theoretical},
\[
H_{\mathrm{Heis}} \;=\; -\sum_{i<j} J_{ij}\,\mathbf{S}_i\!\cdot\!\mathbf{S}_j,
\]
where \(J_{ij}\) is the microscopic exchange coupling—positive \(J_{ij}\) favors ferromagnetic alignment of spins \(\mathbf{S}_i\) and \(\mathbf{S}_j\), while negative \(J_{ij}\) favors antiferromagnetic order.  In the spin–bath analogy for self-attention this exchange tensor is replaced by the learned Query–Key coupling matrix  
\[
W_{\mathrm{eff}} \;=\;\frac{W_Q\,W_K^{\!\top}}{\sqrt{d_{\mathrm{head}}}},
\]
so that the corresponding two-body “attention Hamiltonian” reads  
\[
H^{(0)}(S_j,S_i) \;=\;-\,S_j\,W_{\mathrm{eff}}\,S_i^{\!\top}.
\]
Here, $W_Q$ and $W_K$ are the learned query and key projection matrices for a given attention head, and $d_{\mathrm{head}}$ is the dimensionality of these internal query and key vectors.
In this formulation, each attention head behaves like a spin-coupled system, with the predicted preference for a candidate token determined by the alignment of its key-space projection with an effective context field. The central theoretical object is the so-called Context Vector \(N^{(h)}_0\), which summarizes the aggregated preference of the head "h" from earlier tokens.
Although \(J_{ij}\) originates from orbital overlaps or superexchange processes in a real material and \(W_{\mathrm{eff}}\) is fit to linguistic data during pre-training, both tensors occupy the identical mathematical role of weighting the dot-product of two unit spins.  Consequently, each element \((W_{\mathrm{eff}})_{ij}\) quantifies an effective “semantic exchange” strength in direct analogy to the magnetic exchange \(J_{ij}\), and the eigenmodes of \(W_{\mathrm{eff}}\) reveal principal collective semantics just as diagonalization of \(J_{ij}\) yields magnon eigenmodes in a solid.
The purpose of our work is to test this prediction directly in the pretrained GPT-2 small model~\cite{radford2019language}, moving beyond illustrative examples to a quantitative validation using statistical and causal tools. Our approach isolates the Query–Key interactions from the Value pathway to rigorously assess how well the two-body spin Hamiltonian captures next-token preferences.

\subsection{Theoretical logit difference from a single head's output}
\label{sec:delta_l_theory}

To isolate and quantify the contribution of a single attention head to the final next-token decision, we compute a theoretical logit difference, $\Delta L_{\text{theory}}^{(h)}$, derived from that head's complete output along its Q--K--V--O pathway. This allows us to model the head's influence in the same space in which it is passed to the next stage of the Transformer (either the feed-forward block or the next layer).

We begin with an input prompt comprising $k$ tokens. Their embeddings, drawn from the token embedding matrix $W_E$, form a matrix $\mathbf{X} \in \mathbb{R}^{k \times d_{\text{model}}}$. For a single attention head $h$, the computation involves three learned projection matrices
\[
W_Q^{(h)},\; W_K^{(h)},\; W_V^{(h)} \in \mathbb{R}^{d_{\text{model}} \times d_{\text{head}}}.
\]
The prompt embeddings are projected into the corresponding subspaces as
\[
Q = \mathbf{X} W_Q^{(h)}, \qquad
K = \mathbf{X} W_K^{(h)}, \qquad
V = \mathbf{X} W_V^{(h)}.
\]
For the final token in the prompt (position $k$), its query vector $Q_k$ interacts with all key vectors $K_j$ in the context to form attention scores
\[
\Omega_j = \frac{Q_k \cdot K_j}{\tau}, \qquad \tau = \sqrt{d_{\text{head}}},
\]
which are converted into attention weights $\alpha_j = \mathrm{softmax}(\Omega)_j$.

The primary output of head $h$ at this position is a \emph{context vector} in the head's value space,
\begin{equation}
\mathbf{N}_0^{(h)} = \sum_{j=1}^{k} \alpha_j V_j \in \mathbb{R}^{d_{\text{head}}},
\label{eq:context_vector}
\end{equation}
which aggregates information from all context tokens. To reinject this information into the model's residual stream, the head applies its learned output matrix
\[
W_O^{(h)} \in \mathbb{R}^{d_{\text{model}} \times d_{\text{head}}},
\]
yielding a projected context vector in the model space,
\begin{equation}
\mathbf{N}_{\text{proj}}^{(h)} = W_O^{(h)} \mathbf{N}_0^{(h)} \in \mathbb{R}^{d_{\text{model}}}.
\label{eq:projected_context_vector}
\end{equation}
The vector $\mathbf{N}_{\text{proj}}^{(h)}$ is the complete contribution of head $h$ to the residual stream at the final position.

To obtain a theoretical prediction for this head's next-token preference, we treat $\mathbf{N}_{\text{proj}}^{(h)}$ as an effective field acting on the vocabulary. Let $e_t \in \mathbb{R}^{d_{\text{model}}}$ denote the standard embedding (or unembedding) vector associated with token $t$ (a row of $W_E$ or of the unembedding matrix). We model the theoretical logit contribution of head $h$ to token $t$ as the inner product
\begin{equation}
L_{\text{theory}}^{(h)}(t)
= \mathbf{N}_{\text{proj}}^{(h)} \cdot e_t.
\end{equation}
For two candidate tokens $A$ and $B$, the corresponding theoretical logit difference is therefore
\begin{equation}
\Delta L_{\text{theory}}^{(h)}
= L_{\text{theory}}^{(h)}(A) - L_{\text{theory}}^{(h)}(B)
= \mathbf{N}_{\text{proj}}^{(h)} \cdot (e_A - e_B).
\label{eq:delta_l_theory_final}
\end{equation}
This expression captures the effect of the head's full Q--K--V--O pathway on the logit gap between tokens $A$ and $B$ as seen in the residual stream.

For later geometric analysis it is sometimes convenient to work instead in the head's value space. In that case we define head-specific token vector a for a token, A
\[
S_A^{(h)} := (W_O^{(h)})^{\top} e_A \in \mathbb{R}^{d_{\text{head}}},
\]
so that $L_{\text{theory}}^{(h)}(A) = \mathbf{N}_0^{(h)} \cdot S_A^{(h)}$ and
\begin{equation}
\Delta L_{\text{theory}}^{(h)}
= \mathbf{N}_0^{(h)} \cdot \bigl(S_A^{(h)} - S_B^{(h)}\bigr)
= \mathbf{N}_{\text{proj}}^{(h)} \cdot (e_A - e_B),
\end{equation}
where the last equality follows directly from Eq.~\eqref{eq:projected_context_vector}. In Sec.~\ref{R&D} we exploit this head-space form to visualize the decision geometry of a single head.

\subsection{Logit differences as physically meaningful metrics}

In Transformer models, each token is assigned a scalar logit \(L_i \in \mathbb{R}\), which serves as an unnormalized score reflecting the model’s preference for that token in the given context. These logits are passed through a soft-max function to produce normalized probabilities:
\[
P_i = \frac{e^{L_i/\tau}}{Z}, \qquad Z = \sum_{j \in \mathcal{V}} e^{L_j/\tau},
\]
where \(\mathcal{V}\) denotes the vocabulary and \(\tau = \sqrt{d_{\text{head}}}\) serves as an effective temperature. From a statistical-physics perspective, which is formalized within the framework of energy-based models \cite{lecun2006tutorial} this normalization constant \(Z\) plays the role of a partition function, and each logit corresponds to minus the energy of a token: \(E_i = -L_i\). Hence, the soft-max implements a Boltzmann distribution over candidate tokens.

In this setting, the \textit{logit difference} between two tokens,
\[
\Delta L = L_{\text{good}} - L_{\text{bad}},
\]
directly sets the energy gap \(\Delta E = -\Delta L\) and determines their relative odds via the exponential relation:
\[
\frac{P_{\text{good}}}{P_{\text{bad}}} = e^{\Delta L/\tau}.
\]

This quantity is particularly well-suited for comparing theoretical predictions with actual model behavior for several reasons. First, \(\Delta L\) is invariant under any additive offset to all logits, such as those introduced by residual connections or layer normalizations. Second, it is independent of the absolute value of the partition function \(Z\), which may vary unpredictably across prompts. Third, \(\Delta L\) remains valid under changes in temperature, entering as a linear scaling in \(\exp(\Delta L/\tau)\).

Throughout this study, we adopt \(\Delta L\) as the primary metric of comparison between the theoretical two-body model and GPT-2’s empirical next-token preferences. This choice captures the model’s decision boundary in a form that is physically interpretable, statistically robust, and directly comparable to the Hamiltonian’s energy gap.

\section{Experimental Procedure}
Our multi-stage validation uses this calculation as its core. The procedure is as follows:

\begin{enumerate}
    \item \textbf{Systematic Validation Dataset:} We constructed a dataset of 20 diverse factual recall prompts (see Appendix \ref{app:prompts} for the complete prompt list). For each prompt, we defined a correct "good" token and an incorrect "bad" token.

    \item \textbf{Statistical Correlation Analysis:} We calculated $\Delta L_{\text{theory}}$ for all 144 heads on all 20 prompts. We compared this to the empirical logit difference, $\Delta L_{\text{actual}}$, from the full model's output. We performed a linear regression analysis to identify the most consistently predictive heads and to quantify the statistical significance of their correlation.

    \item \textbf{Causal Ablation:} To test for causality, we used PyTorch hooks to perform head ablation on a representative prompt. By zeroing out a head's contribution during the forward pass, we measured the resulting change in the final model's output, allowing us to distinguish correlational from causal effects.
\end{enumerate}

\section{Results and Discussion.}
\label{R&D}
Our empirical validation of the spin-bath model proceeded in three stages: a systematic, statistical analysis to identify functionally relevant attention heads; a mechanistic visualization of a key head's decision geometry; and a causal intervention to confirm its role.

\subsection{Statistical identification and mechanistic analysis of an antagonistic head}

To move beyond single-prompt anecdotes, we first quantified the predictive power of all 144 attention heads in GPT-2 on a diverse set of 20 factual recall prompts. For each head, we computed the Pearson correlation between its theoretically predicted logit difference, $\Delta L_{\text{theory}}$ (as given by our two--body attention model), and the final logit difference of the full model, $\Delta L_{\text{actual}}$, measured directly from the GPT-2 logits.

\begin{figure}[h]
    \centering
    \includegraphics[width=\columnwidth]{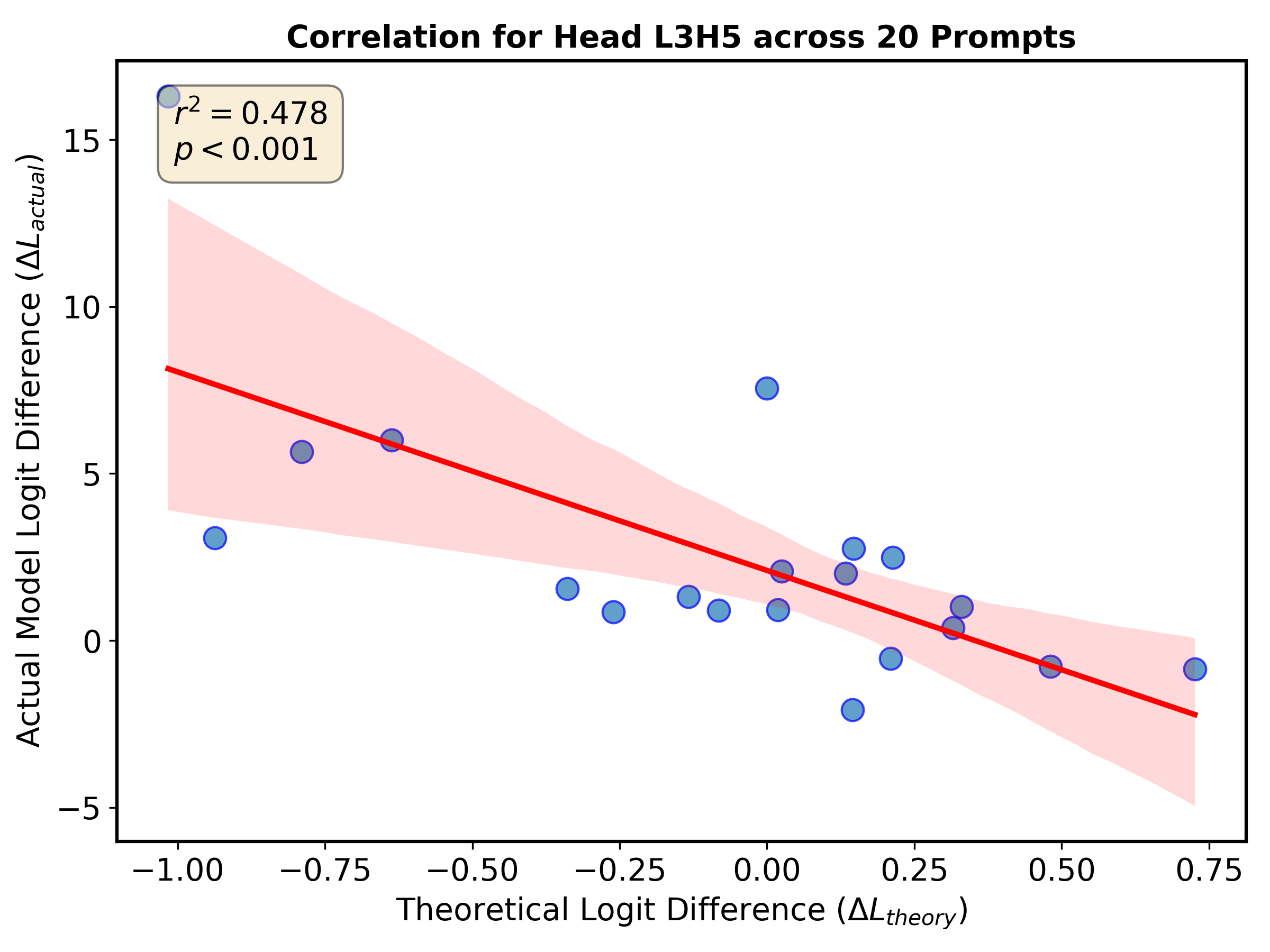} 
    \caption{Statistical Validation of Head L3H5: The theoretical logit difference from head L3H5 ($\Delta L_{\text{theory}}$) is plotted against the final model's actual logit difference ($\Delta L_{\text{actual}}$) for 20 prompts. The strong negative correlation ($r^2=0.478, p<0.001$) demonstrates that this head has a significant and consistently antagonistic function for this task.}
    \label{fig:correlation_scatter}
\end{figure}

This systematic sweep identifies Layer~3, Head~5 (L3H5) as the most systematically predictive component in the sense that its theoretical logit gaps are most strongly correlated (here, anti--correlated) with the model’s empirical logit gaps. As shown in Fig.~\ref{fig:correlation_scatter}, this head exhibits a strong and statistically significant \emph{negative} correlation between $\Delta L_{\text{theory}}$ and $\Delta L_{\text{actual}}$ ($r \approx -0.70$, $p < 0.001$). A linear regression yields a coefficient of determination $r^2 \simeq 0.48$, indicating that the simple two--body model for this single head accounts for nearly half of the variance in the full 12--layer model’s logit gaps on this task set. In other words, L3H5 is not merely active but functionally \emph{antagonistic}: when the model as a whole moves probability mass toward the correct answer, this head’s contribution tends, on average, to push in the opposite direction. This illustrates how the physics--inspired framework can expose non--obvious, countervailing roles played by individual components.

\begin{figure}[t]
    \centering
    \includegraphics[width=\columnwidth]{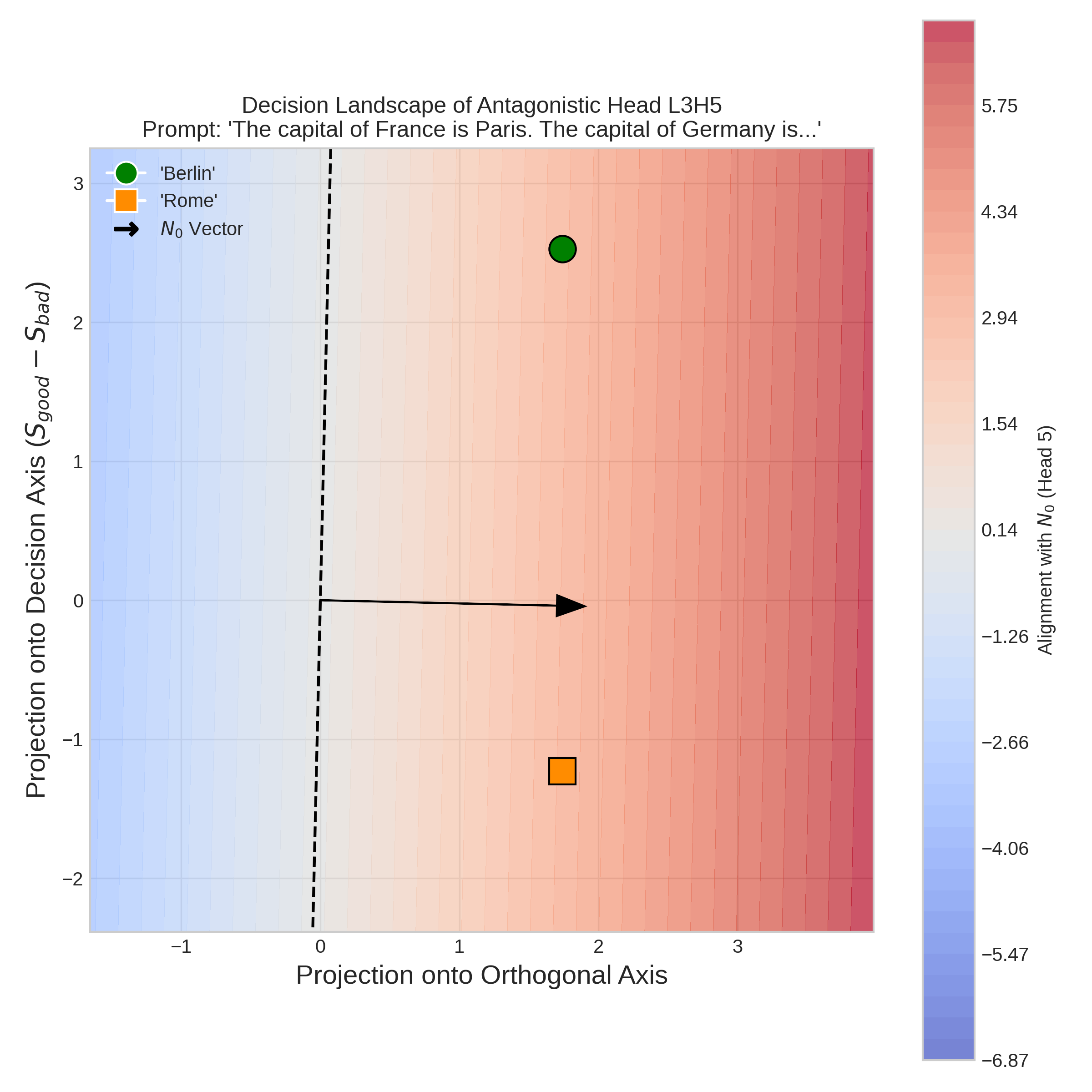}  
    \caption{Decision Landscape of \textit{Head 5 in Layer 3}\\
           Prompt: “The capital of France is Paris. The capital of Germany is…”}
    \label{Fig2}
\end{figure}

To unpack this statistical antagonism mechanistically, Fig.~\ref{Fig2} visualizes the decision geometry of Head~L3H5 in its internal value space for a single prompt. The plane in the figure is spanned by two axes: a vertical \emph{decision axis},
\[
\Delta S = S_{\text{good}} - S_{\text{bad}},
\]
and a horizontal axis orthogonal to it. Here $S_{\text{good}}$ and $S_{\text{bad}}$ denote the embeddings of the candidate tokens in the same head--space used by our theoretical model. The background colormap encodes the head’s internal ``preference field'': each point in the plane corresponds to a hypothetical token embedding $\mathbf{s}$, coloured by the inner product $N_0 \cdot \mathbf{s}$, where $N_0$ is the head’s context vector (plotted as a black arrow). Red regions correspond to directions in which the head makes a \emph{positive} logit contribution, blue regions to directions in which it makes a \emph{negative} contribution.

For the illustrative prompt ``The capital of Germany is~\dots'', the correct continuation ``Berlin'' (green circle) and the incorrect continuation ``Rome'' (orange square) lie on opposite sides of the decision axis. Crucially, in the geometry induced by Head~L3H5, the point corresponding to ``Rome'' lies deeper in the favourable (red) region than the point corresponding to ``Berlin''. The resulting theoretical logit gap for this head alone is
\[
\Delta L_{\text{theory}}
= N_0 \cdot S_{\text{Berlin}} - N_0 \cdot S_{\text{Rome}}
\approx -0.195 < 0,
\]
i.e.\ this head \emph{suppresses} the correct answer and nudges probability toward the incorrect one. The full GPT-2 model, of course, still prefers ``Berlin'' once all heads and layers are combined; Fig.~\ref{Fig2} should therefore be read as a local, mechanistic instance of the antagonism already visible statistically in Fig.~\ref{fig:correlation_scatter}. Together, the two figures show that L3H5 is a highly structured but oppositional component: its internal preference field is coherent and low--dimensional, yet its contribution systematically counters the behaviour of the overall network.
In summary, Layer~3, Head~5 behaves as an \emph{antagonistic} head: along its full Q--K--V--O pathway it produces a negative logit gap for the correct token, $\Delta L_{\text{theory}} < 0$, thereby pushing probability mass away from the model--preferred answer. This logit contribution is a linear functional of the pretrained output matrix $W_O$ via Eq.~(3), so the observed antagonism directly reflects the way $W_O$ projects the head's context vector back into the residual stream.

\subsection{Influence of global generation temperature on output order}
Beyond the effective internal temperature $\tau_{\text{head}} = \sqrt{d_{\text{head}}}$ that scales dot products within individual attention heads (as discussed in Section~II, the final token selection process in LLMs is critically influenced by an externally applied global generation temperature, denoted $T$. This parameter scales the raw output logits, $L_v \in \mathbb{R}$ for each token $v$ in the vocabulary $\mathcal{V}$, before the final softmax layer. The probability $P_v$ of generating token $v$ is then given by the Boltzmann distribution,
\begin{equation}
P_v(T) = \frac{\exp(L_v / T)}{\sum_{j \in \mathcal{V}} \exp(L_j / T)}.
\label{eq:softmax_global_temp}
\end{equation}
This temperature $T$ thereby controls the stochasticity of the output~\cite{guo2017calibration,xie2024calibrating}. To characterize this global effect using physics-inspired metrics, we investigated the model's output distribution as a function of $T$. We performed a temperature sweep from $T=0.1$ to $T=2.5$ for the GPT-2 model across our factual-recall prompt set. Two key order parameters were tracked: the average Shannon entropy of the next-token probability distribution, $S(P(T))$, and the average probability of the most likely next token, $P_{\text{top-1}}(T)$.

The Shannon entropy, a measure of the uncertainty or "disorder" in the probability distribution, was calculated as,
\begin{equation}
S(P(T)) = - \sum_{v \in \mathcal{V}} P_v(T) \log_2 P_v(T),
\label{eq:shannon_entropy}
\end{equation}
where the logarithm is base 2, so entropy is measured in bits. The probability of the top-1 token, $P_{\text{top-1}}(T)$, which quantifies the model's confidence in its most preferred prediction, was determined by:
\begin{equation}
P_{\text{top-1}}(T) = \max_{v \in \mathcal{V}} \{P_v(T)\}.
\label{eq:p_top1}
\end{equation}
These metrics were averaged over all prompts for each value of $T$.
\begin{figure}[t]
    \centering
    \includegraphics[width=\columnwidth]{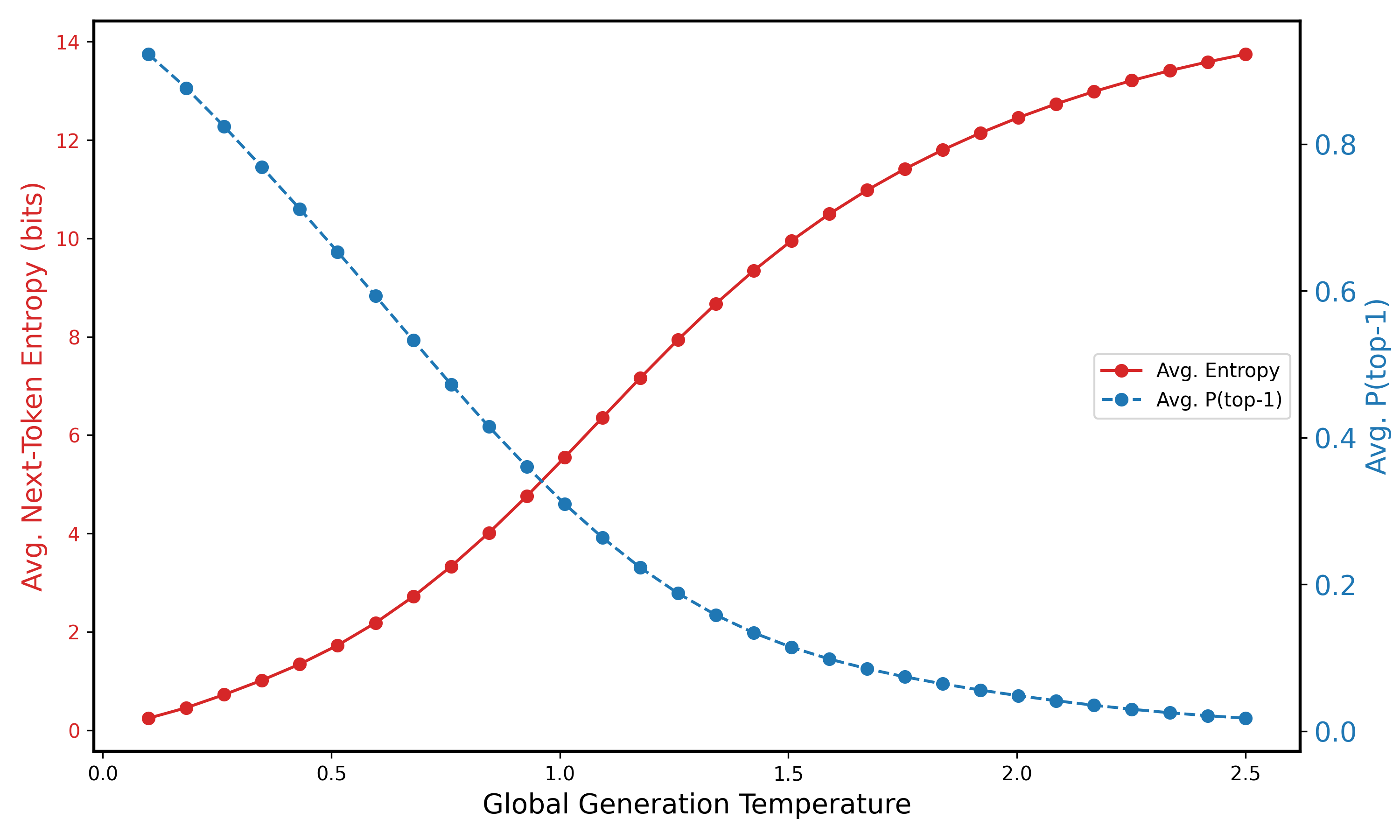} 
    \caption{Order Parameters vs. Global Generation Temperature for GPT-2}
    \label{OP}
\end{figure}
As depicted in Figure~\ref{OP}, the average next-token entropy $S(P(T))$ exhibited a monotonic increase with temperature. At low temperatures ($T \approx 0.1$), entropy was minimal ($\approx 0.2$ bits), signifying a highly deterministic output distribution, analogous to a physical system in a low-energy, ordered state. As $T$ increased, entropy rose towards approximately 14 bits at $T=2.5$, indicating a significantly more stochastic and diverse output, akin to a system exploring a larger volume of its phase space at higher thermal energies. Conversely, $P_{\text{top-1}}(T)$ displayed an inverse relationship with temperature, starting near 0.95 at low $T$ and decreasing to approximately 0.02 at high $T$. This behavior is characteristic of an order parameter, such as magnetization, which diminishes as thermal fluctuations disrupt long-range order in a system. Both parameters showed smooth, continuous changes, indicating a gradual transition from a deterministic to a more random generation regime, rather than a sharp phase transition. A notable crossover region was observed around $T \approx 0.7-1.2$, where the rate of change for both metrics began to moderate. These results empirically validate that the global generation temperature $T$ acts as an effective control mechanism, directly analogous to thermodynamic temperature in physical systems, governing the balance between exploitation of learned patterns and exploration of the output space. This global perspective on output stochasticity complements the head-specific analysis of the spin-bath model.

\subsection{Causal Role of Attention Heads}
Having established a strong correlation, we next performed causal interventions to probe the system's dynamics. We selected a prompt where the baseline model already exhibits an incorrect preference: for ``Lions are carnivores. Cows are...'', the model incorrectly favors `omnivores` over `herbivores`, yielding a negative logit difference ($\Delta L_{\text{actual}} = -0.85$). We then performed head ablation using PyTorch hooks to measure the change in this output~\cite{michel2019sixteen}, a method central to causal analyses of model mechanisms~\cite{meng2022locating}

\begin{figure}[h]
    \centering
    \includegraphics[width=\columnwidth]{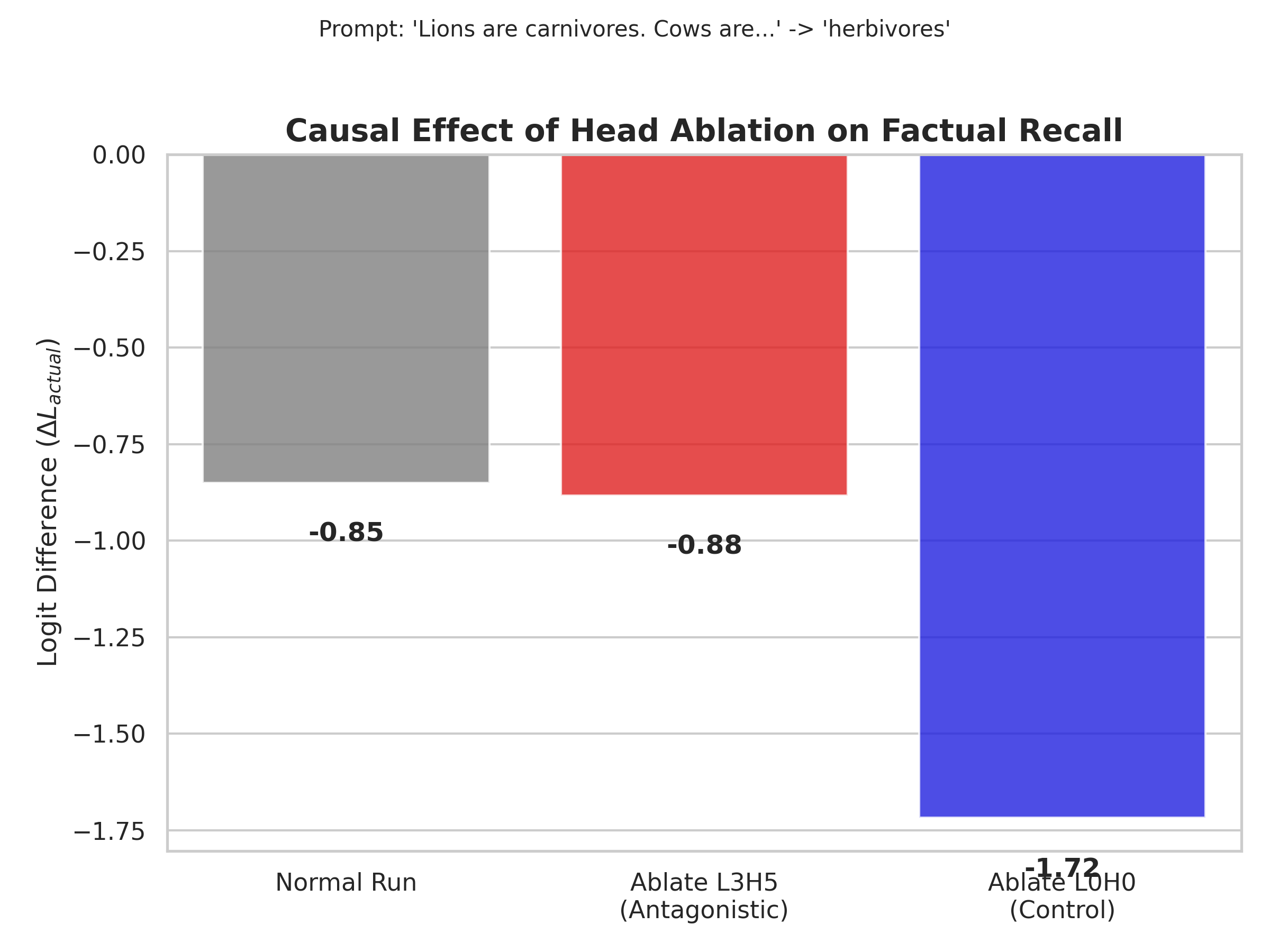} 
    \caption{Causal effect of Head ablation: The bar chart shows the change in the model's output logit difference for a challenging prompt. Ablating the statistically identified antagonistic head L3H5 has a negligible effect. In contrast, ablating a control head, L0H0, significantly worsens the model's output, revealing L0H0's crucial but context-specific positive role.}
    \label{fig:causal_ablation}
\end{figure}

The results, shown in Figure~\ref{fig:causal_ablation}, are revealing. First, ablating the antagonistic head L3H5 had a minimal causal impact on this specific prompt, shifting the logit difference only slightly to -0.88. This suggests that while L3H5 is statistically important across many contexts, its influence can be minor on a case-by-case basis, overshadowed by a coalition of other heads.

More strikingly, when we ablated a control head, L0H0, which had low predictive correlation overall, the model's performance on this prompt degraded significantly, with the logit difference dropping to -1.72. This intervention uncovered a "hidden" function: for this particular context, L0H0 was a key task-aligned component, and its removal exposed the model's underlying difficulty with the prompt.

These causal experiments demonstrate that model behavior is not dictated by single components but emerges from a complex, context-dependent interplay of many heads with both cooperative and antagonistic roles. The final decision is a systemic property, which our methodology successfully probes.

\begin{figure}
    \centering
    \includegraphics[width=\columnwidth]{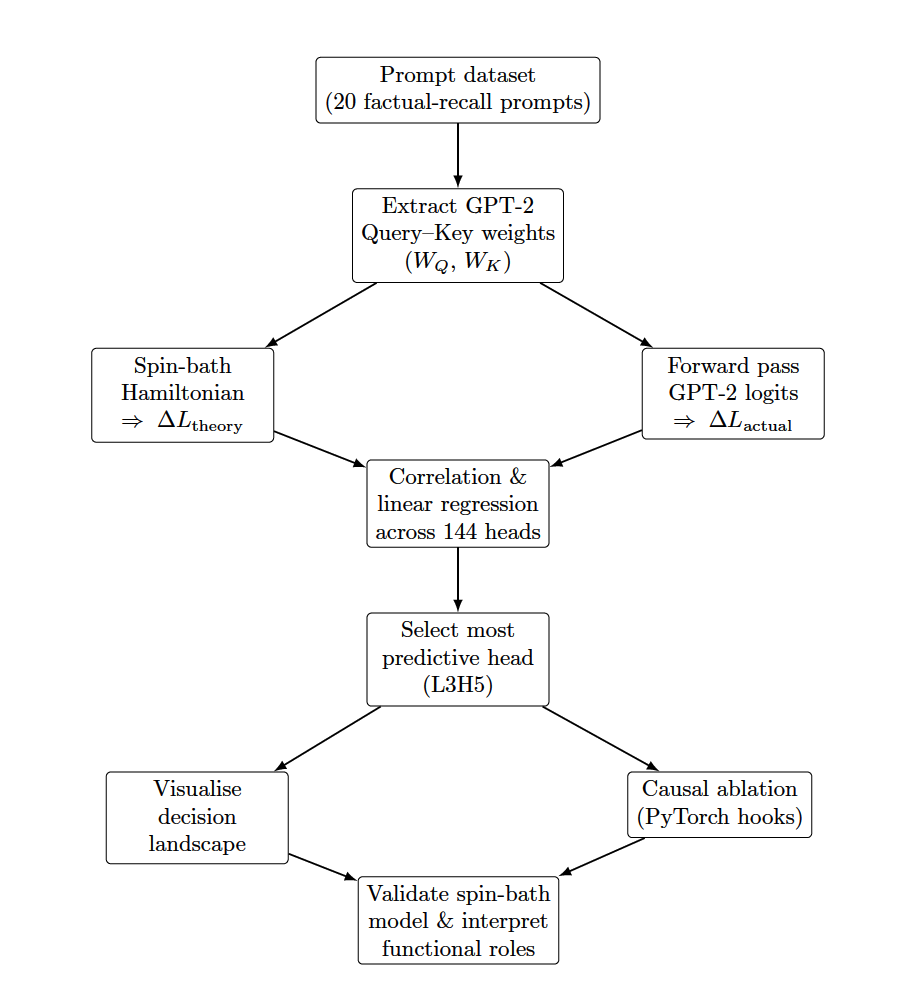} 
    \caption{Workflow used to test the spin-bath model of attention on a pre-trained GPT-2 transformer.}
    \label{Workflow}
\end{figure}
By demonstrating that self-attention in GPT-2 admits an exact mapping onto a two-body spin Hamiltonian—and by validating this mapping both statistically and causally—this study provides the NLP community with a principled, physics-inspired toolkit for interpretability. First, the logit-gap criteria derived from the Hamiltonian give a quantitative lens for ranking head importance that complements purely data-driven metrics. Second, the phase-boundary perspective invites the import of equilibrium and non-equilibrium techniques (e.g. temperature scaling, finite-size scaling) to probe robustness, calibration, and failure modes of attention mechanisms. Finally, by pinpointing both antagonistic and cooperative heads in situ, practitioners gain a mechanistically grounded route to targeted interventions—whether pruning for efficiency, editing to remove undesirable behaviors, or bias-mitigation through controlled perturbations—thus bridging the gap between empirical NLP engineering and theoretical foundations.
\subsection{Towards a three-body extension of the spin-bath model}
\label{sec:three_body_extension}
A natural extension the present two–body Hamiltonian is to incorporate effective three–body couplings between \textit{token spins}. In the spin–bath formulation of attention by Huo and Johnson~\cite{huo2024capturing}, perturbations such as linear bias or positional encoding generate corrections to the context vector that are cubic in the spins and hence mimic an effective three–spin interaction in a constrained space, even though the bare attention energy remains pairwise in form. This suggests that a systematic three–body extension of our model could be constructed by (i) starting from the exact two–body Hamiltonian for a head, (ii) introducing small perturbations that represent either bias, positional structure, or the action of subsequent nonlinear blocks (e.g.\ MLP layers or downstream heads), and (iii) integrating out these perturbations at the level of an effective theory to generate tri–linear terms of the schematic form
\[
H_{\text{Three-body}} \;\sim\; - \sum_{i,j,k} K_{ijk}\,F(\mathbf{S}_i,\mathbf{S}_j,\mathbf{S}_k),
\]
with $K_{ijk}$ an effective three–body kernel and $F$ a suitable symmetric (or structured) trilinear form. Concretely, one could treat the cubic spin contributions that arise in such perturbative expansions as defining $K_{ijk}$ and then ask whether augmenting the two–body Hamiltonian with a small number of structured three–body terms substantially improves the correlation between $\Delta L_{\text{theory}}$ and $\Delta L_{\text{actual}}$. In this view, the current two–body model plays the role of a leading term in a controlled hierarchy, while the three–body corrections encode higher–order dependencies such as multi–token motifs or bias–induced cooperative effects, in line with the conjecture that a generalized three–body attention can capture phenomena that are inaccessible to purely pairwise interactions.

\section{Beyond Brownian Diffusion: Spin‐dynamical generative models}
\label{sec:outlook_spin_dynamics}
\begin{figure}
    \centering
    \includegraphics[width=\columnwidth,height=7cm]{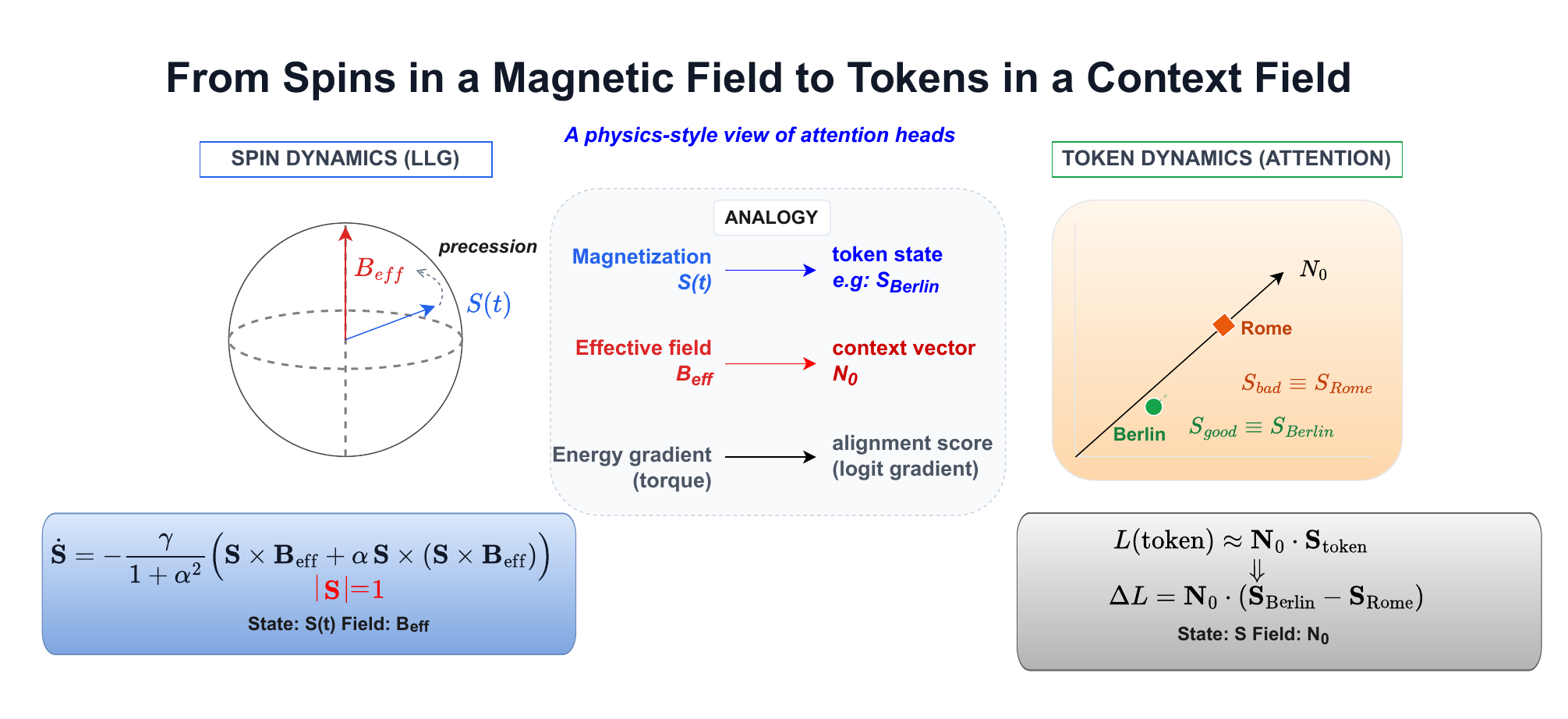}
    \caption{Conceptual analogy between LLG spin dynamics and token scoring by a single transformer attention head. Left: A unit magnetization ($\mathbf{S}(t)\in S^2$) precesses about and relaxes toward an effective field ($\mathbf{B}_{\rm eff}$) according to the Landau–Lifshitz–Gilbert equation (precession + damping). Center: Dictionary: state vector ($\mathbf{S}$) ($\leftrightarrow$) token representation, effective field ($\mathbf{B}_{\rm eff}$) ($\leftrightarrow$) prompt-conditioned context direction ($\mathbf{N}_0$), and damping-driven alignment ($\leftrightarrow$) logit steering. Right: In the value/OV space of Layer 3, Head 5, this head contributes logits approximately ($L(\text{token})\approx \mathbf{N}_0\!\cdot\!\mathbf{S}_{\text{token}}$), giving a logit gap ($\Delta L\equiv L(\text{Berlin})-L(\text{Rome})=\mathbf{N}_0\!\cdot(\mathbf{S}_{\text{Berlin}}-\mathbf{S}_{\text{Rome}})$). Here ($\Delta L<0$), so this head favors the incorrect token Rome over Berlin for the prompt “The capital of Germany is …”, illustrating its antagonistic role within the full GPT-2 model.}

\label{info}
\end{figure}

The static, energetic perspective of the spin-bath model, while powerful for interpreting existing architectures, also suggests a profound extension toward novel, dynamically evolving frameworks. In the following, we can argue that the spin–bath analogy discussed so far is not merely interpretative: it can inspire \emph{dynamical} generative schemes whose core mathematics—spin precession plus viscous damping—surpasses the simplicity of Brownian drift–diffusion in standard score-based models. The situation is further shown graphically in the schematic figure Fig.\ref{info}.

\subsection{From Static Hamiltonian to Dynamical Semantics}

Our two-body attention Hamiltonian,
\[
H^{(0)}(S_j,S_i)\;=\;-\,S_j\,W_{\mathrm{eff}}\,S_i^{\!\top},
\]
has until now offered a purely energetic interpretation of self-attention.  In micromagnetics, however, the same Hamiltonian yields an \emph{effective field},
\[
\mathbf{B}_{\mathrm{eff}} \;=\; -\,\frac{\partial H}{\partial \mathbf{S}},
\]
which drives the Landau–Lifshitz–Gilbert (LLG) dynamics of a spin \cite{gilbert2004phenomenological}.  By analogy, the context‐aggregated value vector
\(\mathbf{N}_{\mathrm{proj}}\) (i.e.\ \(V\,\mathrm{softmax}(QK^{\!\top}/\sqrt{d})\)) plays the role of an \emph{internal field} derived from the prompt.  \(\mathbf{N}_{\mathrm{proj}}\), therefore, can be thought of as an effective "semantic field" on the entire vocabulary of potential next tokens.

\subsection{LLG evolution of a \textit{token spin}}
\label{subsec:token_spin_llg}

We represent each token by a \emph{unit-norm} latent \emph{token spin}
$\mathbf S(t)\in\mathbb S^{d-1}\subset\mathbb R^{d}$.
Its evolution is driven by a context-dependent internal field
$\mathbf N_{\mathrm{proj}}(t)$ and a stochastic (thermal) field $\mathbf h(t)$.
For intuition, in $d=3$ this may be written as a stochastic Landau--Lifshitz--Gilbert (sLLG) equation~\cite{bhattacharjee2014ultrafast}
\begin{equation}
\dot{\mathbf S}= -\gamma\,\mathbf S\times\!\bigl(\mathbf N_{\!\text{proj}}+\mathbf h(t)\bigr)
 -\gamma\frac{\lambda}{1+\lambda^{2}}\,
   \mathbf S\times\!\bigl[\mathbf S\times(\mathbf N_{\!\text{proj}}+\mathbf h(t))\bigr],
\label{sllg}
\end{equation}
with Gaussian white noise $\langle h_\alpha(t)h_\beta(0)\rangle = 2\lambda k_BT\,\delta_{\alpha\beta}\delta(t)$.
The first term generates conservative rotations about $\mathbf N_{\!\text{proj}}$ (precession),
while the double cross product provides dissipative alignment.

For the generative mechanism, the key geometric fact is that the \emph{damping (or double cross) torque}
term is equivalent to a \emph{tangent projection}:
for $\|\mathbf S\|=1$,
\begin{equation}
\mathbf S\times\bigl[\mathbf S\times\mathbf v\bigr]
= (\mathbf S\!\cdot\!\mathbf v)\,\mathbf S-\mathbf v
= -\,\mathbf P_{\mathbf S}\,\mathbf v,
\qquad
\mathbf P_{\mathbf S} \equiv \mathbf I-\mathbf S\mathbf S^{\!\top},
\label{eq:triple_product_projection}
\end{equation}
so the dissipative drift is intrinsically constrained to the tangent plane of $\mathbb S^{d-1}$.
This reformulation remains meaningful in arbitrary dimension $d$, where the projection
$\mathbf P_{\mathbf S}$ (rather than the $3$D cross product) is the natural operator.

In the strong-damping regime ($\lambda\gg 1$) the alignment term dominates and one obtains the
overdamped (Brownian spin-dynamics) limit in which the deterministic part is essentially
$\dot{\mathbf S}\propto \mathbf P_{\mathbf S}\mathbf N_{\!\text{proj}}$.
Setting $\mathbf h\equiv 0$ reduces~\eqref{sllg} to the usual deterministic LLG dynamics~\cite{gilbert2004phenomenological}.

\subsection{Contrast with Brownian Diffusion Models}
Standard diffusion generative models (e.g.\ DDPMs and score-based models) are most often introduced
with an \emph{Euclidean} state variable $\mathbf x\in\mathbb R^{n}$ and a forward \emph{noising} process
that is simple and analytically specified.
In continuous time, this is commonly written as a drift--diffusion SDE of the form~\cite{sohl2015deep,ho2020denoising,sun2022score}
\begin{equation}
d\mathbf x = f(\mathbf x,t)\,dt + g(t)\,d\mathbf W_t,
\label{eq:euclidean_forward_sde}
\end{equation}
together with a learned \emph{denoiser/score} (rather than a learned forward drift).
Concretely, the network is trained to approximate either the score
$s_\theta(\mathbf x,t)\approx\nabla_{\mathbf x}\log p_t(\mathbf x)$
or an equivalent noise/denoising parameterization; this learned object then enters the
\emph{reverse-time} dynamics used for sampling~\cite{sun2022score}.

Our setting differs in two structural ways.

\paragraph{(i) Manifold state space and geometric noise.}
A token is encoded as a unit vector $\mathbf S\in\mathbb S^{d-1}$, not an unconstrained $\mathbf x\in\mathbb R^n$.
Both drift and diffusion must therefore remain tangential to the sphere.
This is enforced by the projector $\mathbf P_{\mathbf S}$ in~\eqref{eq:triple_product_projection},
so norm preservation is built into the continuous-time dynamics.

\paragraph{(ii) Physics-shaped drift and residual learning.}
Rather than learning an arbitrary vector field in ambient space, we start from a structured,
physically motivated drift based on spin dynamics: in the overdamped limit of~\eqref{sllg},
the deterministic flow is proportional to $\mathbf P_{\mathbf S}\mathbf N_{\!\text{proj}}$.
The neural network is then used to supply the \emph{residual score/denoising correction} needed to
invert the forward corruption process on the sphere (cf.\ Section~D).
This provides a strong inductive bias: the model does not need to ``discover'' the entire dynamics
from scratch, but only the part not captured by the analytic spin geometry.

\paragraph{(iii) Irreversibility by design (guidance).}
Finally, our sampler will explicitly re-introduce the context field during denoising.
This produces a \emph{guided} (generally non-reversible) reverse process, which can accelerate mixing and
steer trajectories toward semantically preferred regions of the manifold.
We emphasize this as a \emph{guided} sampler rather than the exact time-reversal of the forward noising SDE.

\subsection{Blueprint of the attention-guided diffusion}
\label{subsec:blueprint}

We call the method \textit{attention-guided diffusion} because an internal
\textit{context field} $\mathbf N_{\!\text{proj}}$ steers an otherwise context-free
diffusion on $\mathbb S^{d-1}$ during generation.

\paragraph*{Forward (noising) process.}
Training begins by corrupting each clean spin $\mathbf S_0\in\mathbb S^{d-1}$ using an
\emph{isotropic spherical diffusion} (a time-rescaled Brownian motion on the sphere)~\cite{de2022riemannian}:
\begin{equation}
d\mathbf S_t \;=\;
\sqrt{\beta(t)}\,\mathbf P_{\mathbf S_t}\!\circ d\mathbf W_t,
\qquad
\mathbf P_{\mathbf S}= \mathbf I-\mathbf S\mathbf S^{\!\top},
\label{eq:OU_forward}
\end{equation}
where $d\mathbf W_t$ is a $d$-dimensional Wiener increment and
$\beta(t)\in[0,\beta_{\max}]$ is a monotonically increasing noise schedule.
Because the increment is projected tangentially (and interpreted in the Stratonovich sense),
the continuous-time process preserves $\|\mathbf S_t\|=1$ and its marginals drift toward the
uniform measure on $\mathbb S^{d-1}$.
This is the manifold analogue of the variance-preserving Gaussian diffusion used in DDPMs:
it progressively erases information and approaches a known prior, but the prior here is
\emph{uniform on the sphere} rather than Gaussian in $\mathbb R^n$.

Since the context field $\mathbf N_{\!\text{proj}}$ is \emph{absent} in the forward phase,
the corruption is Markovian and fully specified.
The model learns the \emph{Riemannian} (tangent) score
\begin{equation}
\mathbf s_\theta(\mathbf S_t,t)\;\approx\;\nabla_{\!\mathbb S}\log q_t(\mathbf S_t),
\qquad
\nabla_{\!\mathbb S}\log q_t(\mathbf S)\equiv \mathbf P_{\mathbf S}\,\nabla_{\!\mathbf S}\log q_t(\mathbf S),
\label{eq:riemannian_score_def}
\end{equation}
from pairs $(\mathbf S_0,\mathbf S_t)$ generated by~\eqref{eq:OU_forward}, where
$q_t(\mathbf S_t)$ denotes the time-marginal density on $\mathbb S^{d-1}$.

\paragraph*{Reverse (denoising / generation) process.}
At sampling time we integrate a \emph{reverse-time score-based SDE} augmented by
\emph{context guidance} (see also Appendix~B):
\begin{equation}
d\mathbf S_t
=
\Bigl[
\eta\,\mathbf P_{\mathbf S_t}\mathbf N_{\!\text{proj}}
-\beta(t)\,\mathbf P_{\mathbf S_t}\,\nabla_{\!\mathbf S}\log q_t^{\mathrm{res}}(\mathbf S_t)
\Bigr]dt
\;+\;
\sqrt{\beta(t)}\,\mathbf P_{\mathbf S_t}\!\circ d\bar{\mathbf W}_t .
\label{eq:LLG_reverse}
\end{equation}
Here $\eta\in[0,1]$ controls the strength of the context field during generation,
$d\bar{\mathbf W}_t$ is reverse-time Wiener noise, and the \emph{residual score}
$\nabla_{\!\mathbf S}\log q_t^{\mathrm{res}}$ is supplied by the neural network.
We stress that the $\eta\,\mathbf P_{\mathbf S}\mathbf N_{\!\text{proj}}$ term makes
\eqref{eq:LLG_reverse} a \emph{guided} reverse-time sampler rather than the exact reverse of the
forward noising process~\eqref{eq:OU_forward}.

Equation~\eqref{eq:LLG_reverse} can be viewed as the overdamped limit of a stochastic LLG:
defining an effective field
$\mathbf H_{\mathrm{eff}}(\mathbf S,t)\equiv
\eta\,\mathbf N_{\!\text{proj}}-\beta(t)\,\nabla_{\!\mathbf S}\log q_t^{\mathrm{res}}$,
the drift is precisely $\mathbf P_{\mathbf S}\mathbf H_{\mathrm{eff}}$,
which is the dissipative (double-cross) LLG torque written in projection form
(cf.\ \eqref{eq:triple_product_projection}).

Starting from a maximally noisy spin $\mathbf S_T$,
one can discretize~\eqref{eq:LLG_reverse} with a Stratonovich-consistent scheme
(e.g.\ stochastic Heun / midpoint); in practice, a final renormalization
$\mathbf S\leftarrow \mathbf S/\|\mathbf S\|$ may be applied to suppress floating-point drift.
As $t\to 0$, noise is annealed and the guided drift steers $\mathbf S_t$ toward regions preferred by
the language model. A nearest-neighbour lookup in embedding space then yields the next token; the token
is appended to the context, $\mathbf N_{\!\text{proj}}$ is recomputed, and the diffusion chain is run
again for the subsequent position.

\begin{figure}
    \centering
    \includegraphics[width=\columnwidth]{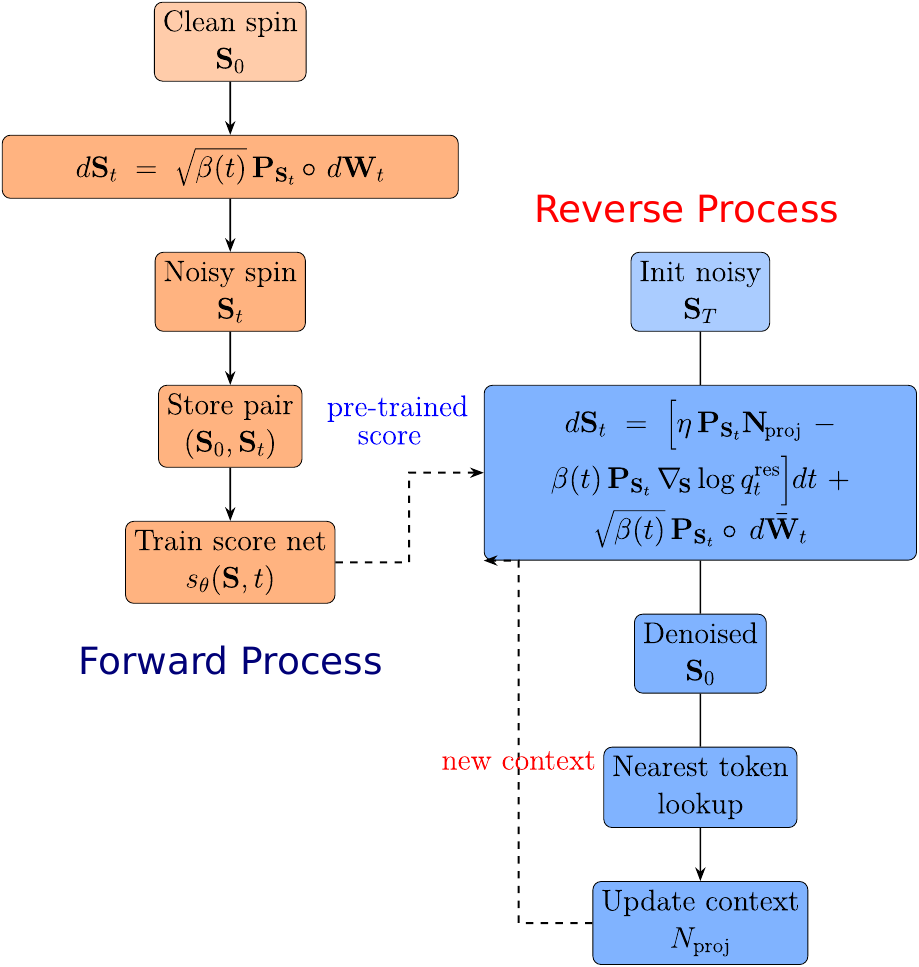}
    \caption{Blueprint of the attention-guided spin-diffusion pipeline.
    Left (forward/noising): each clean spin $\mathbf S_{0}\in\mathbb S^{d-1}$ is corrupted by an isotropic
    spherical diffusion~\eqref{eq:OU_forward} that preserves norm and approaches the uniform prior on the sphere,
    producing a noisy state $\mathbf S_{t}$.
    Right (reverse/generation): starting from a maximally noisy spin $\mathbf S_{T}$, a guided reverse-time
    sampler~\eqref{eq:LLG_reverse}, combining the learned residual score with the context-dependent field
    $\mathbf N_{\!\mathrm{proj}}$ is integrated back to $t=0$ to obtain a denoised spin $\mathbf S_{0}$.}
    \label{Workflow}
\end{figure}

\subsection{Challenges and future directions}

\paragraph{Drift–noise calibration.}
The deterministic drift contains a fixed analytic field $\eta\,\mathbf N_{\mathrm{proj}}$ and a tunable damping coefficient $\lambda$,
while the stochastic kick has variance $\sigma^2(t)$.
Incorrect relative scaling can overdamp the token spin or, conversely, leave it wandering.
A principled schedule for $(\eta,\lambda,\sigma)$ is therefore essential.

\paragraph{Integrator stability.}
Time–stepping a stochastic LLG on the unit hypersphere requires schemes that conserve
$|\mathbf S(t)|=1$ and handle white-noise torques.  
Spin-preserving midpoint or symplectic integrators from micromagnetics, augmented with a projection step, are promising options.

\paragraph{Training objective.}
The score can be split into an analytic part $\eta\,\mathbf N_{\mathrm{proj}}$ and a learned residual $f_\theta$.
The residual is trained with the standard DDPM noise-prediction loss, while $\eta$ can be annealed or learned by meta-gradient.

\paragraph{Empirical validation.}
It remains to be tested whether injecting a physics-informed drift improves perplexity, sample diversity, or controllability compared with a purely learned score field.  
Ablation over $(\eta,\lambda)$ and comparison to Brownian DDPM baselines will provide a decisive benchmark.

\noindent In summary, recasting language generation as spin-dynamical evolution under a self-generated field not only enriches the theoretical landscape—bridging magnetisation dynamics and generative AI—but also offers new inductive biases and control knobs for building next-generation diffusion-style models.

\section{Conclusion}
\label{sec:conclusion}

In this work, we demonstrated that applying the spin–bath Hamiltonian directly to the Query–Key weights of a production-scale Transformer enables quantitatively accurate and falsifiable predictions regarding the behavior of large language models. Our analysis across a diverse set of factual-recall tasks established a strong correlation between analytically computed logit gaps derived from the two-body Hamiltonian and empirically measured gaps, identifying Layer 3 Head 5 as a notably influential unit. Phase-boundary visualizations confirmed this head's contrarian role—actively diverting probability from correct completions—while causal ablations revealed the model’s final outputs as arising from intricate competition among cooperative and antagonistic components. The efficacy of this minimal Hamiltonian highlights that significant aspects of linguistic processing can be effectively modeled using pairwise semantic interactions.

These findings reinforce the conceptual bridge between condensed-matter physics and mechanistic interpretability, translating complex attention mechanisms into accessible frameworks of effective fields, order parameters, and interaction-driven dynamics. This perspective opens possibilities for targeted interventions in model behavior, including precise edits, bias audits, and enhanced robustness. Furthermore, the outlined stochastic Landau–Lifshitz–Gilbert approach offers a physically informed alternative for text generation, incorporating precession and damping dynamics into diffusion-based sampling. Thus, the spin–bath formalism not only elucidates current Transformer operations but also hints at huge possibilities for future progress in physics-inspired language modeling.

\section*{Acknowledgements}
This work was supported by the Korea Institute of Science and Technology (Grant number 2E31851), GKP (Global Knowledge Platform, Grant number 2V6760) project of the Ministry of Science, ICT and Future Planning.
\appendix
\section{Prompt dataset used for causal–intervention tests}
\label{app:prompts}
For completeness we list in Table~\ref{tab:prompts} the 20 hand-crafted
commonsense prompts (and their corresponding \textit{good} and
\textit{bad} continuations) that were used in the to calculate the logit-gap difference and head-ablation analysis in the manuscript.

\begin{table}[h]
\caption{Prompt set employed for causal–intervention experiments.
“Good” is the semantically correct continuation; “Bad” is a plausible
but deliberately incorrect foil chosen to induce a measurable logit
gap.}
\label{tab:prompts}
\begin{ruledtabular}
\begin{tabular}{p{0.43\linewidth}p{0.18\linewidth}p{0.18\linewidth}}
\textbf{Prompt (prefix, including full stop where present)} &
\textbf{Good} & \textbf{Bad}\\ \hline 
The capital of France is Paris. The capital of Germany is & Berlin & Rome\\
The laws of physics are same in all inertial & frames & bodies\\
The author of Hamlet was Shakespeare. The author of The Odyssey was & Homer & Plato\\
An apple is a fruit. A carrot is a & vegetable & mineral\\
The currency of Japan is the Yen. The currency of the United Kingdom is the & Pound & Dollar\\
A dog says woof. A cat says & meow & moo\\
The planet closest to the Sun is Mercury. The second closest is & Venus & Mars\\
To drive a nail, you use a hammer. To tighten a screw, you use a & screwdriver & wrench\\
The opposite of hot is & cold & warm\\
The chemical symbol for water is H\textsubscript{2}O. The chemical symbol for salt is & NaCl & KCl\\
A triangle has three sides. A square has & four & five\\
The first letter of the alphabet is A. The last letter is & Z & Y\\
In the Northern Hemisphere, summer is hot. Winter is & cold & mild\\
Lions are carnivores. Cows are & herbivores & omnivores\\
The main ingredient in bread is flour. The main ingredient in wine is & grapes & apples\\
The color of the sky on a clear day is blue. The color of grass is & green & yellow\\
One dozen is equal to twelve. Two dozen is equal to & twenty‐four & thirty\\
The Pacific is an ocean. The Amazon is a & river & lake\\
A story has a beginning, a middle, and an & end & epilogue\\
To see, you use your eyes. To hear, you use your & ears & nose\\
\end{tabular}
\end{ruledtabular}
\end{table}

\section{Derivation of the drift term}
\label{app:drift}

\vspace{-0.5\baselineskip}
We begin from the analytic Landau--Lifshitz--Gilbert dynamics used in the main text.
For clarity we first write the standard $d=3$ cross-product form,
\begin{equation}
d\mathbf S
\;=\;
\Bigl[
  -\,\gamma\,\mathbf S\times\mathbf N_{\!\mathrm{proj}}
  \;-\;
  \gamma\frac{\lambda}{1+\lambda^{2}}\,
     \mathbf S\times(\mathbf S\times\mathbf N_{\!\mathrm{proj}})
\Bigr]\!dt
\;+\;\text{noise},
\label{eq:B1}
\end{equation}
where $\|\mathbf S\|=1$ and $\mathbf N_{\!\mathrm{proj}}(\mathbf S,t)$ is the context field.
In the implementation, the \emph{geometric} structure is expressed in arbitrary dimension
via tangent projections on $\mathbb S^{d-1}$ (rather than literal cross products).

\paragraph{Projection onto the tangent.}
Using the vector identity
\(
\mathbf a\times(\mathbf b\times\mathbf c)=
\mathbf b(\mathbf a\!\cdot\!\mathbf c)-\mathbf c(\mathbf a\!\cdot\!\mathbf b)
\)
and $\mathbf S\!\cdot\!\mathbf S=1$, the Gilbert (double-cross) torque becomes
\begin{align}
\mathbf S\times(\mathbf S\times\mathbf N_{\!\mathrm{proj}})
&= (\mathbf S\!\cdot\!\mathbf N_{\!\mathrm{proj}})\,\mathbf S-\mathbf N_{\!\mathrm{proj}}
  \;=\; -(\mathbf I-\mathbf S\mathbf S^{\!\top})\,\mathbf N_{\!\mathrm{proj}}
  \;=\;
  -\,\mathbf P_{\mathbf S}\,\mathbf N_{\!\mathrm{proj}},
\end{align}
where
\begin{equation}
\mathbf P_{\mathbf S}\equiv \mathbf I-\mathbf S\mathbf S^{\!\top}
\end{equation}
projects vectors onto the tangent space $T_{\mathbf S}\mathbb S^{d-1}$.
Substituting back into \eqref{eq:B1} shows that the \emph{dissipative} deterministic drift is
\begin{equation}
d\mathbf S
\;=\;
\eta\,\mathbf P_{\mathbf S}\mathbf N_{\!\mathrm{proj}}\,dt
\;+\;\text{noise},
\qquad
\eta\equiv\gamma\frac{\lambda}{1+\lambda^{2}}.
\label{eq:B2}
\end{equation}
In the main construction we work in the \textit{overdamped limit}, i.e.\ we retain the
dissipative tangent drift \eqref{eq:B2} and drop the conservative precession term
$-\gamma\,\mathbf S\times\mathbf N_{\!\mathrm{proj}}$, so that the analytic drift is aligned
with (tangent) score-based denoising.

\paragraph{Embedding in a score-based diffusion on $\mathbb S^{d-1}$.}
We separate the method into a \emph{forward noising process} and a \emph{reverse-time sampler}.

\smallskip
\noindent\textbf{Forward (noising) process:}
To corrupt clean spins $\mathbf S_0\in\mathbb S^{d-1}$ we use an isotropic spherical diffusion
(projected Brownian motion on the sphere) written in Stratonovich form:
\begin{equation}
d\mathbf S_t \;=\; \sqrt{\beta(t)}\,\mathbf P_{\mathbf S_t}\circ d\mathbf W_t,
\qquad \mathbf S_t\in\mathbb S^{d-1}.
\label{eq:Bfwd}
\end{equation}
Because both the drift (here absent) and the noise increment are tangential, \eqref{eq:Bfwd}
preserves $\|\mathbf S_t\|=1$ and relaxes to the uniform prior on $\mathbb S^{d-1}$ as $t$ increases.
(Equivalently, one may write an It\^o form containing the standard curvature drift; the two
representations describe the same spherical diffusion.)

\smallskip
\noindent\textbf{Reverse (denoising) process:}
Let $q_t(\mathbf S)$ denote the time-marginal density of the forward process \eqref{eq:Bfwd}
(with respect to the uniform surface measure).
Score-based diffusion theory implies that denoising is driven by the \emph{Riemannian (tangent) score}
\(
\nabla_{\mathbb S}\log q_t(\mathbf S)
\equiv \mathbf P_{\mathbf S}\nabla_{\!\mathbf S}\log q_t(\mathbf S)
\).
In our construction, we \emph{decompose} this into (i) an analytic, context-dependent tangent drift
$\eta\,\mathbf P_{\mathbf S}\mathbf N_{\!\mathrm{proj}}$ and (ii) a learned \emph{residual} score
that accounts for what is not captured by the analytic field.

Concretely, we employ the following guided reverse-time Stratonovich SDE during generation:
\begin{equation}
d\mathbf S_t =
\Bigl[
  \eta\,\mathbf P_{\mathbf S_t}\mathbf N_{\!\mathrm{proj}}
  \;-\;
  \beta(t)\,\mathbf P_{\mathbf S_t}\,
            \nabla_{\!\mathbf S}\log q_t^{\mathrm{res}}(\mathbf S_t)
\Bigr]dt
\;+\;
\sqrt{\beta(t)}\,\mathbf P_{\mathbf S_t}\!\circ d\bar{\mathbf W}_t.
\label{eq:B3}
\end{equation}
We stress that the inclusion of the context term
$\eta\,\mathbf P_{\mathbf S}\mathbf N_{\!\mathrm{proj}}$ makes \eqref{eq:B3} a
\emph{guided reverse-time sampler} (analogous in spirit to guidance in diffusion models),
rather than the exact mathematical time-reversal of the context-free forward noising
process \eqref{eq:Bfwd}.

We therefore have three contributions:
\begin{itemize}\setlength{\itemsep}{0pt}
\item \textbf{Analytic guidance:} $\eta\,\mathbf P_{\mathbf S}\mathbf N_{\!\mathrm{proj}}$ is a
      physically structured tangent drift (overdamped spin alignment).
\item \textbf{Learned residual:} $-\beta(t)\,\mathbf P_{\mathbf S}\nabla_{\!\mathbf S}\log q_t^{\mathrm{res}}$
      supplies the data-driven correction (a residual Riemannian score) learned offline.
\item \textbf{Geometric noise:} $\sqrt{\beta(t)}\,\mathbf P_{\mathbf S}\circ d\bar{\mathbf W}_t$
      confines stochasticity to $T_{\mathbf S}\mathbb S^{d-1}$, preserving $\|\mathbf S_t\|$ at the SDE level.
\end{itemize}
Equation~\eqref{eq:B3} therefore combines a first-principles tangent drift inspired by micromagnetics
with score-based denoising on a curved manifold, yielding a diffusion sampler that is geometrically
consistent (unit-norm spins) while remaining guided by empirical data statistics.

\section{Relating the Context Field to an effective field in a Heisenberg Magnet}
\label{app:context-heisenberg}

For unit spins $\mathbf S_i\in\mathbb R^{3}$ the classical Heisenberg
Hamiltonian reads
\begin{equation}
H_{\mathrm{Heis}}
\;=\;
-\sum_{i<j}J_{ij}\,\mathbf S_i\!\cdot\!\mathbf S_j
\;=\;
-\tfrac12\sum_{ij}\mathbf S_i^{\!\top}\!J_{ij}\,\mathbf S_j.
\label{eq:C1}
\end{equation}
Taking the partial derivative with respect to $\bf S_i$ gives the
well-known effective field~\cite{bhattacharjee2012theoretical}
\begin{equation}
\mathbf B_{\mathrm{eff}}^{\,i}
\;=\;
-\,\frac{\partial H_{\mathrm{Heis}}}{\partial\mathbf S_i}
\;=\;
\sum_{j}J_{ij}\,\mathbf S_j,
\label{eq:C2}
\end{equation}
i.e.\ a \emph{linear} superposition of neighbouring spins weighted
by the exchange couplings $J_{ij}$.

\paragraph{Bilinear “token–token” Hamiltonian for one head.}
In a single self-attention head we may coarsen the transformer
weights into an anisotropic exchange tensor $W_{\mathrm{eff}}\in\mathbb
R^{d\times d}$ and write the pair energy as
\begin{equation}
H^{(0)}(\mathbf S_j,\mathbf S_i)
\;=\;
-\,\mathbf S_j\,W_{\mathrm{eff}}\,\mathbf S_i^{\!\top}.
\label{eq:C3}
\end{equation}
A direct derivative, in complete analogy with~\eqref{eq:C2}, yields the
\emph{token-level effective field}
\begin{equation}
\mathbf N_{\!i}^{(0)}
\;=\;
-\,\frac{\partial H^{(0)}}{\partial\mathbf S_i}
\;=\;
W_{\mathrm{eff}}^{\!\top}\,\mathbf S_j.
\label{eq:C4}
\end{equation}
Thus the scalar exchange $J_{ij}$ of the Heisenberg model is replaced by
an anisotropic matrix coupling that multiplies the partner spin.

\paragraph{From one head to many heads/layers.}
For a genuine transformer layer each head $h$ has its own query, key,
value and output matrices
$\{W_Q^{(h)},W_K^{(h)},W_V^{(h)},W_O^{(h)}\}$.
With soft-attention weights
\(
\alpha_{ij}^{(h)}\!\propto\!
\exp\!\bigl[(\mathbf S_iW_Q^{(h)})
           (\mathbf S_jW_K^{(h)})^{\top}/\sqrt d\bigr],
\)
the value-aggregated context vector for token $i$ reads
\begin{equation}
\mathbf N_{0}^{(h)}(i)
\;=\;\sum_{j}\alpha_{ij}^{(h)}
       \bigl(\mathbf S_j W_V^{(h)}\bigr).
\label{eq:C5}
\end{equation}
Projecting back to the model space with $W_O^{(h)}$ gives the
\emph{context field} employed in the main text,
\begin{equation}
\boxed{\,%
\mathbf N_{\mathrm{proj}}^{(h)}(i)
\;=\;
\mathbf N_{0}^{(h)}(i)\;
\bigl[W_O^{(h)}\bigr]^{\top}
\,}.
\label{eq:C6}
\end{equation}

\section{Details of the systematic search for the key head}
\label{app:systematic_search}

We carried out a systematic, head-by-head screening of GPT-2 (12 layers $\times$ 12 heads $= 144$ attention heads) on a small but diverse set of 20 factual–recall prompts. Each prompt was paired with a ground-truth ``good'' token and a plausible ``bad'' distractor (e.g., ``Berlin'' vs.\ ``Rome'' for the capital of Germany). For every head $(\ell,h)$ and every prompt, we computed two quantities:

\begin{itemize}
    \item A \emph{theoretical} logit margin $\Delta L_{\text{theory}}^{(\ell,h)}$, obtained from our head-level Q--K--V--O model (Sec.~\ref{sec:delta_l_theory}).
    \item The corresponding \emph{empirical} logit margin $\Delta L_{\text{actual}}$, measured directly from the pretrained GPT-2 logits at the last position.
\end{itemize}

For a given prompt, let $S \in \mathbb{R}^{k \times d_{\text{model}}}$ denote the matrix of token embeddings for the $k$-token prefix, and let $e_{\text{good}}, e_{\text{bad}} \in \mathbb{R}^{d_{\text{model}}}$ be the model-space embeddings of the good and bad candidate tokens. For head $(\ell,h)$ we extract the per-head projections
\[
W_Q^{(\ell,h)},\; W_K^{(\ell,h)},\; W_V^{(\ell,h)} \in \mathbb{R}^{d_{\text{model}} \times d_{\text{head}}}, 
\qquad
W_O^{(\ell,h)} \in \mathbb{R}^{d_{\text{model}} \times d_{\text{head}}}
\]
from the corresponding GPT-2 attention block. The usual attention computation at the final position $k$ reads
\[
Q = S W_Q^{(\ell,h)}, \quad
K = S W_K^{(\ell,h)}, \quad
V = S W_V^{(\ell,h)},
\]
\[
\Omega = \frac{Q K^\top}{\sqrt{d_{\text{head}}}}, \qquad
\alpha^{(\ell,h)} = \mathrm{softmax}\bigl(\Omega_{k,\cdot}\bigr),
\]
and the head’s value-space context vector is
\begin{equation}
N_0^{(\ell,h)} = \sum_{j=1}^{k} \alpha^{(\ell,h)}_j V_j
\;=\;
\bigl(\alpha^{(\ell,h)}\bigr)^\top \bigl(S W_V^{(\ell,h)}\bigr)
\in \mathbb{R}^{d_{\text{head}}}.
\end{equation}
This is projected back into the residual stream via
\begin{equation}
N_{\text{proj}}^{(\ell,h)} = W_O^{(\ell,h)} N_0^{(\ell,h)} \in \mathbb{R}^{d_{\text{model}}}.
\end{equation}
Following Sec.~\ref{sec:delta_l_theory}, the theoretical contribution of head $(\ell,h)$ to the logit gap between the two candidate tokens is then
\begin{equation}
\Delta L_{\text{theory}}^{(\ell,h)}
= N_{\text{proj}}^{(\ell,h)} \cdot \bigl(e_{\text{good}} - e_{\text{bad}}\bigr).
\label{eq:app_deltaL_theory}
\end{equation}
The corresponding full-model margin for the same prompt is defined as
\begin{equation}
\Delta L_{\text{actual}}
= \mathrm{logit}_{\text{good}} - \mathrm{logit}_{\text{bad}},
\end{equation}
where the logits are taken from the final GPT-2 output at position $k$ when the model is run on the unmodified prompt.

For each head $(\ell,h)$ we thus obtain 20 pairs
\[
\bigl(\Delta L_{\text{theory}}^{(\ell,h)}(\text{prompt } i),\;
      \Delta L_{\text{actual}}(\text{prompt } i)\bigr), 
\qquad i = 1,\dots,20,
\]
and compute the Pearson correlation coefficient $r^{(\ell,h)}$ between the theoretical and empirical margins across prompts. Heads are then ranked by $|r^{(\ell,h)}|$ as a measure of how strongly and systematically the simple two-body theory tracks the behaviour of the full 12-layer model on this task set.

Layer~3, Head~5 (L3H5) emerges as the most strongly correlated head in absolute value, with
\[
r^{(3,5)} = -0.691, \qquad r^{(3,5)2} = 0.478, \qquad
p = 7.38 \times 10^{-4} \quad (n = 20).
\]
The negative sign indicates that L3H5 is \emph{antagonistic}: when the full GPT-2 model increases the logit margin in favour of the correct token, the linear prediction from this head tends, on average, to move in the opposite direction. For example, the top three heads by $|r^{(\ell,h)}|$ that we calculated were:
\begin{align*}
\begin{array}{l}
\text{L3H5 }(-0.691), 
\text{L0H2 }(+0.547), 
\text{L0H7 }(+0.542)
\end{array}
\end{align*}
This ranking motivated our choice of L3H5 as a ``key head'' for detailed mechanistic analysis in the main text.

\clearpage
\newpage
\bibliography{Ref}
\bibliographystyle{apsrev4-2}

\end{document}